\documentclass[aps,prb,graphicx,twocolumn]{revtex4-1}

\usepackage{graphicx}% Include figure files
\usepackage{dcolumn}% Align table columns on decimal point
\usepackage{bm}% bold math
\usepackage[utf8]{inputenc}
\usepackage[T1]{fontenc}
\usepackage{mathptmx}
\usepackage{amsmath}

\begin{document}

%\preprint{}

% \title{Optimized Si/SiO$_2$ nanobeam cavity for linear and non-linear applications} %Title of paper
%\title{A Silica encapsulated photonic crystal nanobeam cavity optimized for high-Q and small footprint} %Title of paper
\title{Global optimization of an encapsulated Si/SiO$_2$ L3 cavity for ultra-high quality factor} %Title of paper



\author{J.P. Vasco}
\email[]{juan.vasco@epfl.ch}
\affiliation{Institute of Physics, \'Ecole Polytechnique F\'ed\'erale de Lausanne (EPFL), CH-1015 Lausanne, Switzerland}

\author{V. Savona}
%\email[]{vincenzo.savona@epfl.ch}
\affiliation{Institute of Physics, \'Ecole Polytechnique F\'ed\'erale de Lausanne (EPFL), CH-1015 Lausanne, Switzerland}


%\date{\today}

\begin{abstract}
We optimize a silica-encapsulated silicon  L3 photonic crystal cavity for ultra-high quality factor by means of a global optimization strategy, where the closest holes surrounding the cavity are varied to minimize out-of-plane losses. We find an optimal value of $Q_c=4.33\times10^7$, thus setting a new record for encapsulated low-index-contrast photonic crystal cavities. We also address the effects of structural imperfections on our optimal cavity design and predict an averaged $Q_c$ in the 2 million regime for state-of-the-art silicon fabrication tolerances.
\end{abstract}


\maketitle %\maketitle must follow title, authors, abstract and \pacs


Photonic crystal (PC) slab cavities have been focus of intense research during the last two decades due to their unique properties to efficiently confine light at length scales close to the diffraction limit, and extremely low loss rates \cite{vahala,noda}. These features have allowed to study a wide variety of classical and quantum phenomena, where the linear and non-linear interactions between light and matter are effectively enhanced in the cavity region \cite{deppe,vuckovic3,imamoglu,noda3,vuckovic4,vuckovic2,vuckovic6,wong,nomura,notomi3,galli,vuckovic5,imamoglu2,notomi4,imamoglu3,faolain,noda4,momchil3}. Broadly speaking, the strength of this enhancement grows with the local density of electromagnetic states, which is proportional to the quality factor of the cavity mode $Q_c$, and inversely proportional to its mode volume $V$ \cite{hughes,hughes2,vasco4}. Hence, massive efforts have been directed toward the optimization of these figures of merit in order to reach the desired functionality of the photonic device \cite{chalcraft,momchil4,noda5,momchil6,noda6,momchil8}. Particularly, silicon-based cavities have attracted very much attention because of their natural compatibility with CMOS technologies and negligible material losses at telecom wavelengths, allowing the integration with optoelectronic devices in a single chip \cite{wang}. While free-standing silicon PC slabs offer an excellent platform to build ultralow loss cavities \cite{noda2,momchil6,dario5}, silica (SiO$_2$) encapsulation improves the mechanical stability and thermal dissipation of the system \cite{raineri}, while mitigating additional loss channels coming from the etching of air holes in the silicon \cite{borselli}. Nevertheless, high quality factors are challenging in such encapsulated structures given the low refractive index contrast between the two materials.

In this letter, we employ a global optimization approach to maximize the quality factor of a Si/SiO$_2$ L3 PC cavity. We find an optimal quality factor of $Q_c=4.33\times10^7$ which corresponds to the largest value achieved for low-index-contrast PC cavities. Our results set a new record for the L3 paradigm and open the way to a new class of highly efficient optical devices for linear and non-linear applications in classical and quantum photonics.

\begin{figure}[h!]
\includegraphics[width=0.5\textwidth]{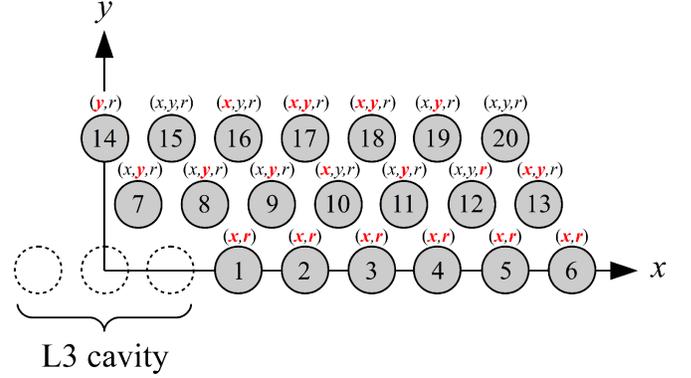}
\caption{Schematic representation of the closest holes surrounding the L3 cavity, in the first quadrant, which are considered in the global optimization procedure. Mirror symmetry with respect to $x=0$ and $y=0$ is assumed, thus setting a total of 53 optimization parameters. Nevertheless, only the ones highlighted in red are found to be the most relevant to increase the fundamental mode quality factor.}\label{fig1}
\end{figure}

\begin{table*}[t!]
\caption{\label{tab1} Summary of the main linear and non-linear figures of merit of the non-optimized and optimized Si/SiO$_2$ L3 cavities.}
\begin{ruledtabular}
\begin{tabular}{ccccccc}
Si/SiO$_2$ -- L3 cavity & $f$ (Thz) & $Q_c$ & $V_l$ $(\lambda/n_{\rm Si})^3$ & $V_{nl}$ $(\lambda/n_{\rm Si})^3$ & $Q_c/V_l$ $(n_{\rm Si}/\lambda)^3$ & $Q_c^2/V^2_{nl}$ $(n_{\rm Si}/\lambda)^6$ \\
\hline
Non-optimized & 195.2 & $1.33\times10^3$ & 0.67 & 3.25 & $1.99\times10^3$ & $1.68\times10^{5}$ \\
Optimized & 191.2 & $4.33\times10^7$ & 1.75 & 7.47 & $2.47\times10^7$ & $3.36\times10^{13}$\\
\end{tabular}
\end{ruledtabular}
\end{table*}

\begin{figure}[h!]
\includegraphics[width=0.48\textwidth]{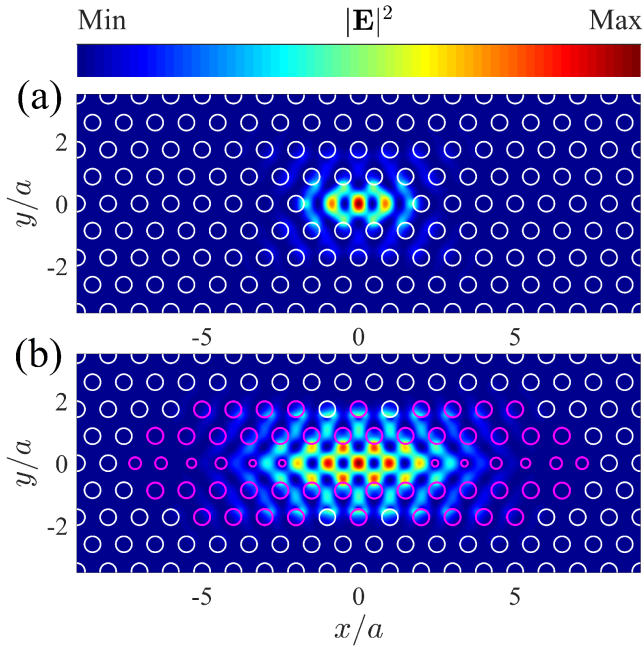}
\caption{(a) Near-field intensity distribution of the non-optimized L3 fundamental mode cavity. (b) same as (a) for the optimized L3 cavity, where the holes which are actually considered in the optimization are represented by magenta circles.}\label{fig2}
\end{figure}

We consider a silica-encapsulated silicon PC slab with a hexagonal lattice of holes of radii $r=100$~nm, lattice parameter $a=390$~nm and thickness $d=220$~nm. A L3 cavity is introduced by removing three holes along the $\Gamma K$ direction of the lattice. In order to optimize the quality factor $Q_c$ of its fundamental mode, we adopt a global optimization approach in which only the closest holes surrounding the cavity are varied, in size $r\rightarrow r+dr$ and position $(x,y)\rightarrow (x+dx,y+dy)$, to reduce out-of-plane losses.  This technique has been extremely successful during the last few years to reach record theoretical and experimental quality factors for a wide variety of different materials and cavity geometries \cite{momchil3,momchil4,momchil5,momchil7,flayac3,momchil6,vasco3}. Specifically, we employ the particles swarm (PS) algorithm to achieve this goal with $Q_c$ as the objective function and the guided mode expansion method (GME) \cite{gme} as the main PC solver. We show in Fig.~\ref{fig1} the schematic representation of the holes to be considered in the optimization procedure, where mirror symmetry with respect to the planes $x=0$ and $y=0$ is assumed. Notice that in such a way we end up with a total of 53 optimization parameters, however, after 1400 iterations of the PS algorithm we have noticed that the most relevant parameters for increasing $Q_c$ are those highlighted in red in Fig.~\ref{fig1}. This preliminary analysis allowed us to reduce the dimension of the optimization parameter space from 53 to 27, thus effectively decreasing the number of function evaluations required by the algorithm to converge. We summarize in Table~\ref{tab1} our final results where 
\begin{equation}\label{Vl}
 V_l=\frac{\int \epsilon(\mathbf{r})|\mathbf{E}(\mathbf{r})|^2d\mathbf{r}}{\mbox{Max}\left\{\epsilon(\mathbf{r})|\mathbf{E}(\mathbf{r})|^2 \right\}},
\end{equation}
is the linear mode volume and
\begin{equation}\label{Vnl}
 V_{nl}=\frac{\left[\int \epsilon(\mathbf{r})|\mathbf{E}(\mathbf{r})|^2d\mathbf{r}\right]^2}{\int \epsilon^2(\mathbf{r})|\mathbf{E}(\mathbf{r})|^4d\mathbf{r}},
\end{equation}
is the non-linear one \cite{painter}, with $\epsilon(\mathbf{r})$ representing the dielectric function of the system and $\mathbf{E}(\mathbf{r})$ the electric field of the cavity mode. A global maximum of $Q_c=4.33\times10^7$ (computed with FDTD \cite{lumerical}) is found after 806200 function evaluations, leading to an improvement of four orders of magnitude with respect to the non-optimized cavity. This theoretical quality factor is the largest reported for silica-encapsulated PC cavities so far \cite{quan1,bazin,vasco3}, setting a new record for ultra-high $Q$ cavities in low-index-contrast PCs. It is important to notice that, different from previous optimizations of the L3 cavity \cite{momchil4}, $Q_c$ is optimized at the expense of the linear and non-linear mode volumes, nevertheless, we still get extremely large enhancement factors $Q_c/V_l$ and $Q_c^2/V^2_{nl}$ which are in the $10^7$ and $10^{13}$ regimes, respectively. The increase of the mode volume is clearly seen in the Fig.~\ref{fig2}, where we plot the near-field intensity distribution of the fundamental cavity mode in the middle of the slab, for the non-optimized cavity, Fig.~\ref{fig2}(a), and the optimized one, Fig.~\ref{fig2}(b). The holes which are actually varied are represented by magenta circles in Fig.~\ref{fig2}(b). The optimal parameters of the cavity as well as the far-filed projection of the near-field components are reported in the Appendices~\ref{appL3silica1} and \ref{appL3silica2}, respectively.

The same optimization strategy can be directly applied to the air-bridge silicon L3 cavity within the same parameter space of dimension 27. For this configuration, we have obtained an FDTD quality factor $Q_c=1.91\times10^8$ which is around 20 times larger than the previous record obtained with deep neural networks \cite{noda7}. While our optimization requires a much larger number of evaluations to find the global maximum of the objective function, it clearly shows that there is still considerable room for further improvement of these figures of merit when increasing the size of the optimization parameter space. Detailed results for the Si/Air L3 cavity are given in the Appendix~\ref{L3air}.

\begin{figure}[h!]
\includegraphics[width=0.5\textwidth]{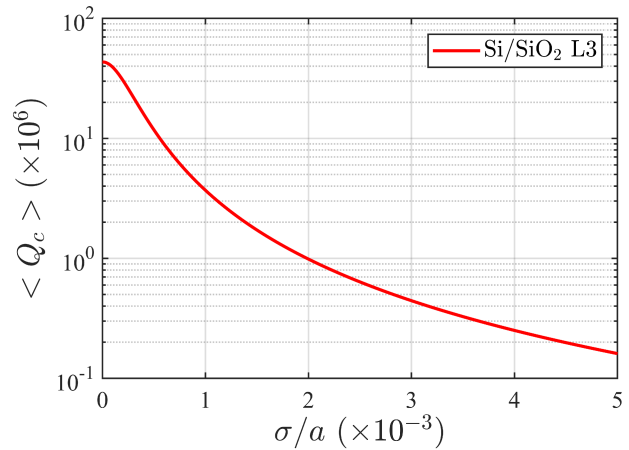}
\caption{Averaged $Q_c$, computed over 100 independent disorder realizations of the optimal cavity, as a function of the disorder parameter $\sigma$. }\label{fig3}
\end{figure}

Since any realistic sample is always subject to a small amount of intrinsic disorder, coming from unavoidable imperfections introduced during the fabrication stage, we model such effect by considering random Gaussian fluctuations in all hole positions and sizes of our PC, where the standard deviation of the Gaussian probability distribution $\sigma$ is taken as the disorder parameter \cite{dario3,momchil2,vasco1}. Results of this analysis are shown in Fig.~\ref{fig3}, where the averaged cavity quality factor $\langle Q_c\rangle$, computed over 100 independent disorder realization of the system, is plotted as a function of $\sigma$. Typical tolerances in silicon state-of-the-art fabrication techniques range between $\sigma=0.001a$ and $\sigma=0.002a$ \cite{noda2,mohamed2}, leading to an averaged $Q_c$ in the 2 million regime, which still correspond to a record figure of merit for silica encapsulated PC structures.

In conclusion, we have optimized a silica-encapsulated silicon L3 cavity by means of a global optimization strategy, where the closest holes surrounding the cavity are varied to decrease out-of-plane losses. We have found a record value of $Q_c=4.33\times10^7$ which is around four times larger than the previous best obtained for Si/SiO$_2$ structures, achieved in nanobeam geometries. To better relate our optimal design to realistic samples, we have also studied the effects of intrinsic disorder and found that when considering typical tolerances in modern fabrication techniques, the averaged quality factor of the optimized cavity remains in the 2 million regime, corresponding to an outstanding result given the low-index contrast of the Si/SiO$_2$ configuration. Apart from setting a new record for the L3 cavity, our results clearly show that there is still a vast room for further improvement of different figures of merit in photonics when increasing the size of the optimization space, and open the way to a new class of optimized designs in low-index-contrast materias, such as AlN, GaN or  Si$_3$N$_4$, holding great promise for nonlinear optical enhancement, sensing, and solid-state quantum optics.

\newpage

\onecolumngrid

\appendix

\section{Optimal parameters of the encapsulated L3 cavity}\label{appL3silica1}
The optimal 27 parameters of the Si/SiO$_2$ L3 cavity with an FDTD quality factor of $Q_c=4.33\times10^7$ are reported in Table~\ref{taba1}.

\begin{table*}[h!]
\caption{\label{taba1} Optimal parameters of the Si/SiO$_2$ L3 cavity with $Q_c=4.33\times10^7$}
\begin{ruledtabular}
\begin{tabular}{ccccccccccc}
Parameter$/$Hole & 1 & 2 & 3 & 4 & 5 & 6 & 7 & 8 & 9 & 10  \\
\hline
$dx$ & 0.4407 & 0.3817 & 0.3936 & 0.3352 & 0.3097 & 0.1385 & $\times$ & $\times$ & $\times$ & 0.0010 \\
$dy$ & $\times$ & $\times$ & $\times$ & $\times$ & $\times$ & $\times$ & 0.0109 & 0.0107 & 0.0082 & $\times$ \\
$dr$ & -0.1500 & -0.1500 & -0.0772 & -0.1075 & -0.0690 & -0.0672 & $\times$ & $\times$ & $\times$ & $\times$  \\
\toprule
Parameter$/$Hole & 11 & 12 & 13 & 14 & 15 & 16 & 17 & 18 & 19 & 20  \\
\hline
$dx$ & $\times$  & $\times$  & -0.0044 & $\times$  & $\times$  & 0.0010 & 0.0018 & 0.0017 & $\times$  & $\times$  \\
$dy$ & 0.0027 & $\times$  & 0.0121 & -0.0010 & $\times$  & $\times$  & -0.0044 & -0.0081 & -0.0071 & $\times$  \\
$dr$ & $\times$  & 0.0001 & $\times$  & $\times$  & $\times$  & $\times$  & $\times$  & $\times$  & $\times$  & $\times$  \\
\end{tabular}
\end{ruledtabular}
\end{table*}

\section{Fourier transform of near-field components}\label{appL3silica2}
\begin{figure}[h!]
\includegraphics[width=0.9\textwidth]{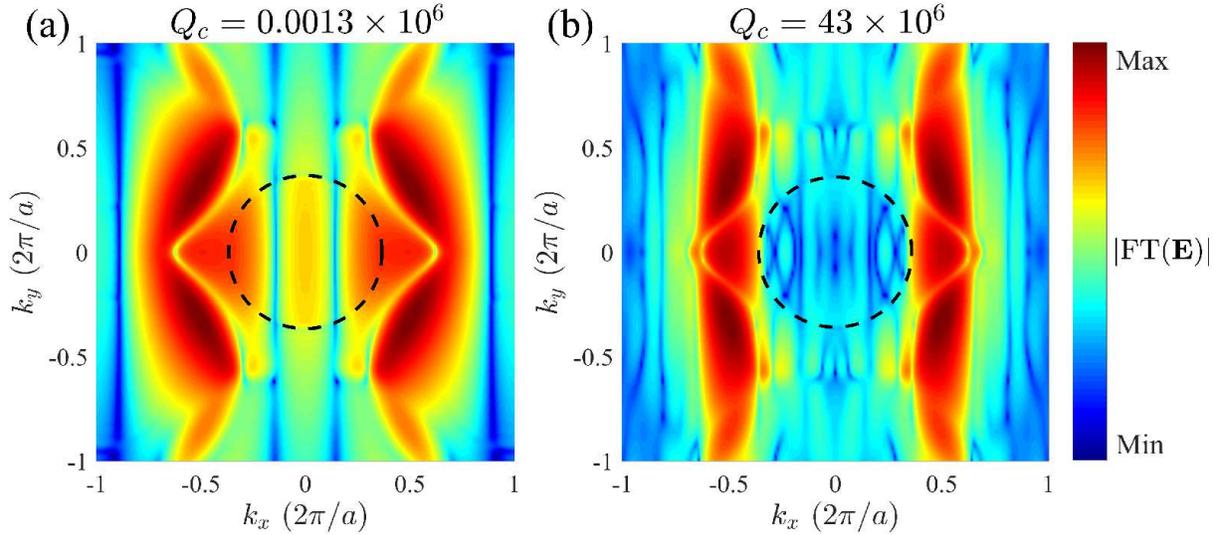}
\caption{(a) Far-field of the non-optimized cavity. (b) same as (a) for the optimized design. The dashed circle represents the region where the cavity frequency crosses the light-line of the dielectric slab.}\label{figs1}
\end{figure}

The far-field projections of the mode components for the non-optimized and optimized cavities, are shown in Figs.~\ref{figs1}(a)~and~\ref{figs1}(b), respectively, in log scale. This projection is obtained through the Fourier transform of the near-field components \cite{vuckovic}, recorded in a $xy$ plane localized at $90$~nm above the photonic crystal surface. The dashed circle represents the region where the cavity frequency crosses the light-line. The strong reduction of the field components inside the light cone (or equivalently, above the light-line) is clearly seen for the optimized design.

\newpage

\section{Results for the Si/Air L3 cavity}\label{L3air}
The Si/Air (air-bridge) L3 cavity is considered in a silicon PC with a hexagonal lattice of holes with radii $r=100$~nm, lattice parameter $a=400$~nm and slab thickness $d=220$~nm. The PS optimization is carried out by considering the same 27 parameters of the Si/SiO$_2$ case. 

\subsection{Optimal figures of merit}
We show in Table~\ref{taba2} the linear and non-linear figures of merit of both, non-optimized and optimized designs. The quality factor is improved by four orders of magnitude with a final FDTD value of $Q_c=1.91\times10^8$, which is around 20 times larger than the previous best, for the silicon L3 cavity, obtained with deep learning optimization techniques \cite{noda7}. 

\begin{table*}[h!]
\caption{\label{taba2} Linear and non-linear figures of merit for the non-optimized and optimized Si/Air L3 cavities.}
\begin{ruledtabular}
\begin{tabular}{ccccccc}
Si/Air -- L3 cavity & $f$ (Thz) & $Q_c$ & $V_l$ $(\lambda/n_{\rm Si})^3$ & $V_{nl}$ $(\lambda/n_{\rm Si})^3$ & $Q_c/V_l$ $(n_{\rm Si}/\lambda)^3$ & $Q_c^2/V^2_{nl}$ $(n_{\rm Si}/\lambda)^6$\\ 
\hline
Non-optimized & 196.3 & $6.53\times10^3$ & 0.59 & 2.48 & $1.10\times10^4$ & $6.94\times10^{6}$ \\
Optimized & 193.6 & $1.91\times10^8$ & 1.07 & 4.30 & $1.78\times10^8$ & $1.97\times10^{15}$\\
\end{tabular}
\end{ruledtabular}
\end{table*}

\subsection{Optimal parameters}
The optimal parameters for the Si/Air L3 cavity with $Q_c=1.91\times10^8$ are reported in Table~\ref{taba3}

\begin{table*}[h!]
\caption{\label{taba3} Optimal parameters of the Si/Air L3 cavity with $Q_c=1.91\times10^8$}
\begin{ruledtabular}
\begin{tabular}{ccccccccccc}
Parameter$/$Hole & 1 & 2 & 3 & 4 & 5 & 6 & 7 & 8 & 9 & 10  \\
\hline
$dx$ & 0.3800 & 0.2954 & 0.2000 & 0.4032 & 0.2360 & 0.0475 & $\times$ & $\times$ & $\times$ & -0.0179 \\
$dy$ & $\times$ & $\times$ & $\times$ & $\times$ & $\times$ & $\times$ & -0.0232 & -0.0157 & 0.0028 & $\times$ \\
$dr$ & -0.0445 & -0.0174 & 0.0033 & -0.0433 & -0.1500 & -0.0805 & $\times$ & $\times$ & $\times$ & $\times$  \\
\toprule
Parameter$/$Hole & 11 & 12 & 13 & 14 & 15 & 16 & 17 & 18 & 19 & 20  \\
\hline
$dx$ & $\times$  & $\times$  & -0.0341 & $\times$  & $\times$  & 0.0040 & -0.0001 & -0.0059 & $\times$  & $\times$  \\
$dy$ & -0.0600 & $\times$  & -0.0141 & -0.0078 & $\times$  & $\times$  & 0.0083 & -0.0114 & -0.0307 & $\times$  \\
$dr$ & $\times$  & -0.0427 & $\times$  & $\times$  & $\times$  & $\times$  & $\times$  & $\times$  & $\times$  & $\times$  \\
\end{tabular}
\end{ruledtabular}
\end{table*}

\newpage

\subsection{Disorder analysis}

\begin{figure}[h!]
\includegraphics[width=0.6\textwidth]{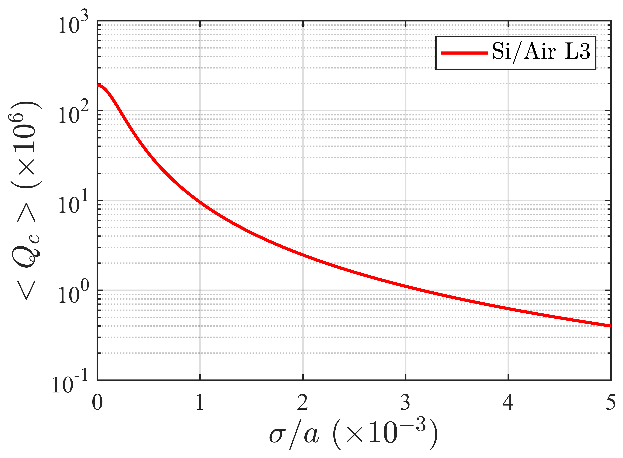}
\caption{Averaged $Q_c$, computed over 100 independent disorder realizations of the optimal Si/Air cavity, as a function of the disorder parameter $\sigma$.}\label{figs2}
\end{figure}

Figure~\ref{figs2} show the disorder analysis for the optimal air-bridge L3 cavity. An averaged $Q_c$ in the 4 million regime is predicted for typical tolerances, ranging between $\sigma=0.001a$ and $\sigma=0.002a$, in silicon fabrication techniques \cite{noda2,mohamed2}.

%\bibliography{bibliography}


\end{document}